\documentclass[%
 reprint,
nofootinbib,
 amsmath,amssymb,
 aps,
]{revtex4-1}

\usepackage{subcaption}
\usepackage{graphicx}
\usepackage{dcolumn}
\usepackage{bm}

\begin{document}

\title{Longitudinal negative magnetoresistance and  magneto-transport phenomena  in conventional and topological conductors}

\author{A. V. Andreev}
\affiliation{Department of Physics, University of Washington, Seattle WA 98195, USA}

\author{B. Z. Spivak }
\affiliation{Department of Physics, University of Washington, Seattle WA 98195, USA}

\date{\today}

\begin{abstract}
Recently a large negative longitudinal (parallel to the magnetic field) magnetoresistance was observed in Weyl and Dirac semimetals. It is believed to be related to the chiral anomaly associated with topological electron band structure of these materials. We show that in a certain range of parameters such  a phenomenon can also exist in conventional centrosymmetric and time reversal conductors, lacking topological protection of the electron spectrum and the chiral anomaly.  We also discuss the magnetic field enhancement of the longitudinal  components of the thermal conductivity and thermoelectric tensors.
\end{abstract}

\maketitle

One can distinguish  two types of magnetoresistance depending on the mutual orientation of the current and the magnetic field: transverse  and longitudinal.
 If the magnetic field is sufficiently small, the magnetoresistance can be  described by the quasiclassical Boltzmann kinetic equation (see for example \cite{AbrikosovBook,LifshitsBook,azbel1973conduction,pippard1989magnetoresistance,kaganov1979electron}).
 A change in the transverse resistance due to a magnetic field can be related to the fact that electrons experience Lorentz force in that direction.  Since there is no Lorentz force in the direction parallel to the magnetic field, the origin of the longitudinal magnetoresistance  is more complicated.   Moreover, the longitudinal magnetoresistance is absent  in the approximation of a spherical Fermi surface, and in the relaxation time approximation \cite{AbrikosovBook}.
 Although no theorem was proven, so far all results based on  the conventional Boltzmann kinetic equation correspond to positive longitudinal magnetoresistance, (see for example Refs.~\onlinecite{kaganov1979electron,Pal2010necessary} and references therein).  Nielsen and Ninomiya~\cite{Nielsen1983}  suggested a chiral anomaly-related~\cite{Adler1969axial,Bell1969pcac} mechanism of negative longitudinal magnetoresistance (NLMR) in materials with massless Dirac and Weyl electronic spectra, which recently attracted great theoretical interest
\cite{WanVishwanath2011,BurkovBalents2011,vafek2014dirac,hirschberger2016chiral}.
The calculations of Ref.~\onlinecite{Nielsen1983} were done in the ultra-quantum limit at zero temperature, and in the case where the chemical potential is at the Dirac point.  However,
in most of existing Dirac and Weyl semimetals the chemical potential is located away from the Dirac points.  In this case a quasiclassical description of the chiral anomaly-related NLMR was developed in Refs.~\onlinecite{SonSpivak2013,spivak2016magnetotransport}. It was shown that the existence of strong NLMR requires a large ratio between the chirality and transport relaxation times.
Recently large NLMR was observed both in Weyl and in Dirac materials (see for example  Refs.~\onlinecite{li2016chiral,xiong2015evidence, yang2015chiral, li2015giant,hirschberger2016chiral, Cano2017chiral}).

In Weyl  semimetals the gapless character of the electron spectrum  is protected by topology. In Dirac metals the massless Dirac points are protected only by the crystalline symmetry.
Therefore a small lattice distortion of a Dirac semimetal can open a gap in the electronic spectrum making it non-topological. Below we consider magnetoresistance in Dirac-type materials in which the electron spectrum is either massless or has a small gap.
Existence of a small gap in a Dirac semimetal was reported already in the first observation of NLMR in these materials \cite{li2016chiral}. Furthermore, NLMR was observed in Weyl materials in which the  Weyl valleys merge into a single electron pocket with zero net topological charge~\cite{Shekhar2015}.
This implies that existence of massless Dirac points in the spectrum, their topological protection and the chiral anomaly are not necessary ingredients  of large NLMR.

In this article we show that negative contributions to the longitudinal  magnetoresistance and other longitudinal magnetotransport phenomena  exist even in conventional centrosymmetric and time-reversal symmetric semiconductors and metals.  However, for this contribution to dominate the effect a certain hierarchy of relaxation times should take place.

To illustrate the origin of the effect we consider a model \cite{Kane1957band}  where the energy gap $E_{g}$ between  between the conduction and the valence bands is significantly smaller than the energy separation from other bands, and the external potential $V({\bm r})$ is smooth on the interatomic scale. In this case the electron dynamics may be described  by the Dirac Hamiltonian (for a recent review see Ref.~\onlinecite{Zawadzki2017semirelativity})
\begin{equation}
\label{Dirac}
\hat{H}=u\,  \bm{p}\cdot \boldsymbol{\sigma}\,  \tau_3+E_g \sigma_{1}+V (\bm{r}).
\end{equation}
Here, $\bm{p}= - i \hbar \boldsymbol{\nabla}-\frac{e}{c}\bm{A} (\bm{r})$ (with $\bm{A} (\bm{r})$ being the vector potential) is the kinematic momentum, $E_g$ is half the band gap, and  $\sigma_i$ and $\tau_i$ denote the Pauli matrices that act in the spin and chirality subspaces respectively.

We focus on  the typical situation in which the electron chemical potential $\mu$ is larger than the gap $E_{g}$.  In this regime electron transport my be described by two equivalent approaches. The first  one is based on the quasiclassical kinetic equation,  while in the second one the free electron motion is described in terms of the Landau levels.
Here we will use the latter approach.  In a uniform magnetic field ${\bf B}$  directed along $z$-direction the energy spectrum of Eq.~\eqref{Dirac} has the form (see for example \cite{Berestetskii1982quantum})
\begin{equation}\label{DiracSpectrum}
 \epsilon^{2}_{n}(p_{z})=E_g^2 + u^2 p_{z}^{2}+  \frac{u^2 \hbar^2}{l_B^2} \, \left( 2n+1 +\sigma \right).
 \end{equation}
Here $p_z$ is the electron momentum along the magnetic field, $l_B = \sqrt{\hbar c/|e B|}$ is the magnetic length, $n=0,1,2...$ labels Landau levels, and $\sigma= \pm 1$ is a spin index.

At $E_{g}=0$ the Hamiltonian \eqref{Dirac} decouples into a sum of Weyl Hamiltonians describing right- and left-handed chiral fermions. As a result,  the electronic states can be classified by chirality ($R$ and $L$), $
\tau_3 \Psi_R=\Psi_R, \quad  \tau_3 \Psi_L= - \Psi_L $. All Landau levels except the lowest one ($n=0, \sigma= -1$) are double degenerate. The electron states in these levels consist of opposite chirality pairs. The spectrum of the lowest Landau level  consists of two nondegenerate linear branches, $\epsilon_0= \pm u p_z$, formed by the states with opposite chirality. As a result, in the the presence of electric field the system exhibits the chiral anomaly  \cite{Nielsen1983,Adler1969axial,Bell1969pcac}.  Acceleration of the electrons by the electric field creates population imbalance of electrons with different chirality. Since the  Hamiltonian Eq.~\eqref{Dirac} decouples into a pair of chiral ($L$ and $R$) Weyl Hamiltonians in the presence of an arbitrary potential $V(\mathbf{r})$, scattering by disorder does not relax the chirality imbalance. Therefore, even at full momentum relaxation of the electron distribution with a given chirality there is an a finite electric current proportional to the chirality imbalance (chiral magnetic effect)~\cite{Vilenkin,Fukushima2008chiral}. In this approximation the electrical conductivity is infinite.

\begin{figure}[t!]
\centering
    \includegraphics[width=1.0\columnwidth]{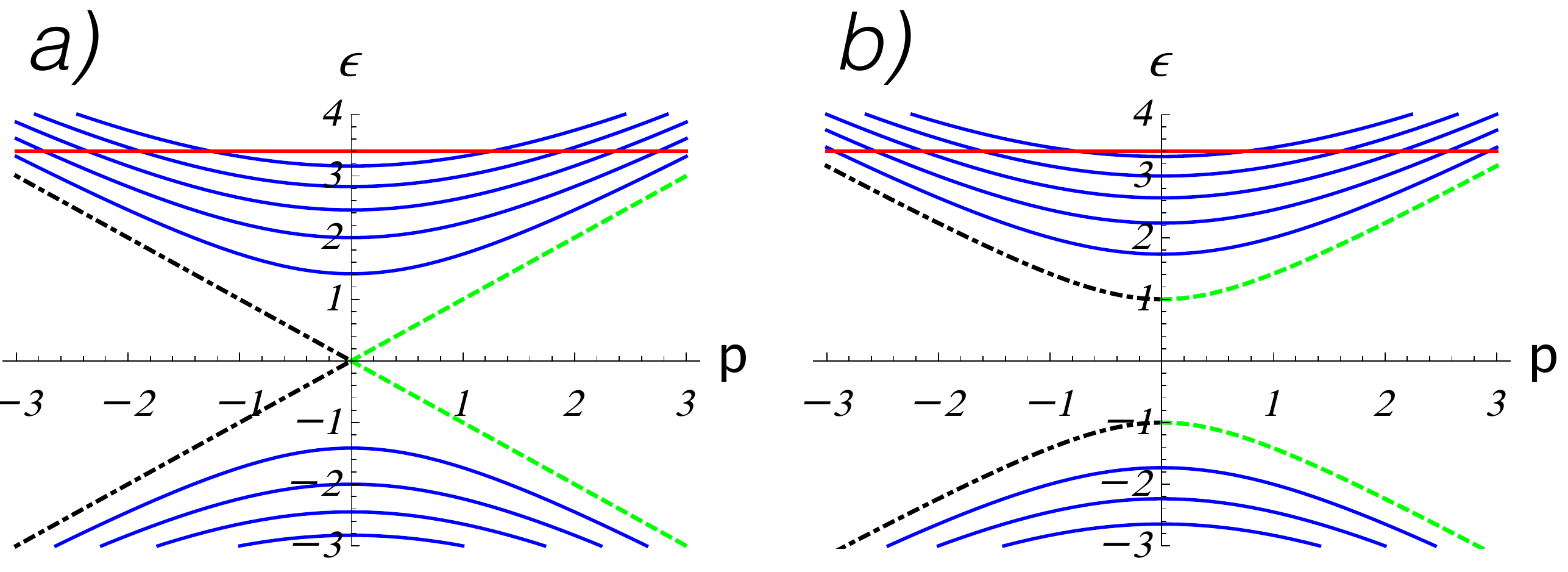}
\caption{\footnotesize Landau level spectrum of the Dirac equation in the gapless case, $E_g=0$ [panel $a)$ ] and gapped case [panel $b)$]. All Landau levels except the lowest one are degenerate in helicity and are shown by solid blue lines. The lowest Landau level is nondegenerate. The helicity of electronic states in it is indicated by the line style: Positive helicity states are shown by the green dashed line, and negative helicity states -- by the black dash-dotted line. The red horizontal line indicates the location of the chemical potential.}
\label{fig:LL_ab}
\end{figure}

At $E_g\neq 0$ chirality is no longer conserved as the second term in Eq.~\eqref{Dirac}, $E_g\tau_1$,  couples the Weyl fermions with opposite chirality. However, since the helicity operator $\hat{\alpha}={\bf p}\cdot \boldsymbol{\sigma}/p$ commutes with the free electron Hamiltonian,  the states of the free electron motion may be classified by the helicity eigenvalues, $\alpha = \pm 1$ (at $E_g=0$ helicity of free electron states coincides with chirality up to the sign of the electron energy). The helicity content of Landau level states is shown in Fig.~\ref{fig:LL_ab}.
 The states in the doubly degenerate Landau levels come in opposite helicity pairs, while the helicity of states in the  non-degenerate  lowest Landau level is given by $\alpha = \mathrm{sign} (p_z)$.

Although at $E_g \neq 0$ there is no chiral anomaly, the mechanism of longitudinal magnetoresistance  is quite similar to that due to the chiral anomaly. Namely,  the acceleration of electrons by the electric field directed along $\mathbf{B}$ produces helicity imbalance. The helicity imbalance in turn produces an electric current even at full momentum relaxation within a population of electrons with the same helicity.  In contrast to chirality, helicity is not conserved by disorder scattering. Nevertheless, it the Fermi energy $E_F$ strongly exceeds the gap $E_g$ the helicity relaxation rate  is parametrically small.
As is shown in Appendix~\ref{sec:helicity_relaxation},  in this regime the helicity relaxation time $\tau_h (\varepsilon)$ may be expressed in terms of the transport mean free time $\tau_{tr} (\varepsilon)$ as
\begin{equation}\label{timehierarchi}
\frac{\tau_h (\varepsilon)}{ \tau_{tr} (\varepsilon)} =  \xi \,  \frac{4 \varepsilon^{2}}{ E_{g}^{2}}  \, \gg \, 1.
\end{equation}
Here $\xi$ is a numerical coefficient of order unity which depends on the angulal dependence of the impurity scattering cross-section. In the Born approximation it is given by Eq.~\eqref{eq:Upsilon}.
Below we develop a theory of electron magnetotransport phenomena  in the leading approximation in $\tau_{tr}/\tau_h$.

In the regime $\tau_{tr}/\tau_h \ll 1$,  during a short time $\tau_{tr}$ the
electron distribution becomes isotropic in momentum and becomes dependent only on the electron energy $\varepsilon$ and helicity $\alpha =\pm 1$, i. e. takes the form  $n_{\alpha}(\varepsilon)$.
In the leading approximation in the parameter $\tau_{tr}/\tau_h$ equations describing electronic transport have the form
\begin{eqnarray}
\partial_t n_{\alpha}(\varepsilon) & = & -\frac{\boldsymbol{\nabla}\cdot \boldsymbol{j}_\alpha(\varepsilon)}{\nu_\alpha (\varepsilon)}
-\frac{k_{\alpha} (\varepsilon)}{\nu_\alpha(\varepsilon)}\frac{e^{2} \boldsymbol{E} \cdot \boldsymbol{B} }{h^2 c} \, \partial_\varepsilon n_{\alpha}^{(0)}  (\varepsilon)    \nonumber \\
&& - \frac{n_\alpha (\varepsilon)  - n_{-\alpha} (\varepsilon)}{\tau_h (\varepsilon)} + I^{in}_\alpha \{n_\alpha (\varepsilon) \}, \label{DiffEq}
\end{eqnarray}
Here $h=2 \pi \hbar$,
$\nu_\alpha(\epsilon)$ is the density of states with helicity $\alpha$, and $I^{in}_\alpha \{n_\alpha (\varepsilon) \}$ denotes the collision integral due to inelastic electron-electron and electron-phonon scattering processes, and we expressed the collision integral due to impurity scattering in terms of the helicity relaxation time, see Eq.~\eqref{eq:collision_integral_helicity}.
The parameter $k_{\alpha}$
describes the  flux of helicity imbalance created by acceleration of electrons in the lowest Landau level by the electric field. It may be expressed in terms of the dispersion relation of this Landau level and is given by
\begin{equation}\label{eq:k_def}
  k_\alpha (\varepsilon) =  \alpha \sqrt{1 - \frac{  E_g^2}{\varepsilon^2} }.
\end{equation}
Finally,
\begin{equation}
{\bf j}_{\alpha}(\varepsilon)=
\frac{ek_{\alpha}n _{\alpha}(\varepsilon)}{h^2 c}\boldsymbol{B}  \label{Bcurrent}
\end{equation}
denotes the density of particle current with helicity $\alpha$  per unit energy. The electric current ${\bf j}$ and the heat flux ${\bf j}_q$ may be expressed  as
\begin{align}\label{current2}
& {\bf j}= e \sum_\alpha
\int d \varepsilon \, {\bf j}_{\alpha}(\varepsilon), \quad
{\bf j}_q = \sum_{\alpha}\int d \varepsilon (\varepsilon - \mu)\,  {\bf j}_{\alpha}(\varepsilon),
\end{align}
where $\mu $ is the chemical potential. Note that  in the limit $\tau_{tr} (\varepsilon)/\tau_h (\varepsilon) \to 0$ both
the net current ${\bf j}_\alpha(\varepsilon)$ and the helicity pumping are associated  with only the lowest Landau  level.
This is the reason why there is a density of states in the denominator in  the first term in the right hand side  of   Eq.~\eqref{DiffEq}.

To leading order in $E_g/\varepsilon \ll 1$  the parameter $k_\alpha (\varepsilon)$ in Eq.~\eqref{eq:k_def} is given by $k_\alpha (\varepsilon) = \alpha =\pm 1$. In this case Eqs. ~\eqref{DiffEq}-\eqref{current2} coincide with those obtained in Refs.~\onlinecite{SonSpivak2013,spivak2016magnetotransport} for Weyl semimetals with topologically protected gapless electron spectrum. In
 Weyl semimetals $k_\alpha = \pm 1$  is given by the quantized monopole charge of the Berry curvature flux and
 Eq.~\eqref{DiffEq} describes the chiral anomaly.  The above consideration shows that both generation of helicity imbalance due to acceleration of electrons by the electric field described by Eq.~\eqref{DiffEq},  and the current proportional to helicity imbalance, Eq.~\eqref{Bcurrent},  exist in generic conductors with no topological protection of the electron spectrum.

Below we  discuss  longitudinal  magnetotransport phenomena: NLMR, enhancement  of thermal conductivity and the thermoelectric effect by a magnetic field. Generally speaking, linear response  phenomena are characterized by tensor transport coefficients. Equations \eqref{DiffEq}-\eqref{current2}, on the other hand, describe only the ``anomalous" contributions to the transport coefficients which affect only the $zz$ components of the tensors. Here  $\hat z$ is the direction of the magnetic field.

Using  Eq.~\eqref{DiffEq} and assuming that $n_\alpha (\varepsilon)=n_F(\varepsilon)+\delta n_\alpha (\varepsilon)$, where $n_F (\varepsilon)=[e^{(\varepsilon-\mu)/T(\bf r) } +1]^{-1}$ is the locally-equilibrium Fermi distribution function, we get
\begin{eqnarray}
\label{eq:delta_n}
I^{in}_\alpha \{n_\alpha (\varepsilon) \}
  & = & -  \frac{\delta n_\alpha (\varepsilon)  -\delta  n_{-\alpha} (\varepsilon)}{\tau_h (\varepsilon)} +  \nonumber \\
&& \frac{ ek_{\alpha} \left(  e \boldsymbol{E} - \frac{\varepsilon - \mu}{T} \,
\boldsymbol{\nabla} T
 \right) \cdot \boldsymbol{B} }{\nu_\alpha (\varepsilon) h^2 c}
\,  \partial_\varepsilon n_F  (\varepsilon) .
\end{eqnarray}
Note that although both terms in the right hand side are odd in  $k_\alpha$ their effect on the nonequilibrium distribution function is drastically different. Only the first term creates the helicity imbalance whereas the second term creates an energy imbalance between the electron populations with different helicities. The inelastic collisions relax this energy imbalance but not the helicity imbalance. As a result the nonequilibrium distribution function may be written in the form
\begin{equation}
\label{eq:delta_n_result}
\delta n_\alpha (\varepsilon)  =
 \frac{ ek_{\alpha} \left( \tau_{eff}\frac{\varepsilon - \mu}{T} \,
\boldsymbol{\nabla} T  -  \tau_h (\varepsilon)  e \boldsymbol{E}
 \right) \cdot \boldsymbol{B} }{2 \nu_\alpha (\varepsilon) h^2 c}
\, \frac{ d n_F  (\varepsilon)}{d \varepsilon} .
\end{equation}
Here $1/\tau_{eff}$  is the effective rate of energy transfer between the electron populations with opposite helicity. Treating the inelastic collision integral in the relaxation time approximation we may express it as
\begin{equation}
 1/\tau_{eff}=1/\tau_h+1/\tau_{\epsilon},
\end{equation}
where $1/\tau_{\epsilon}$ is the inelastic relaxation rate.

Substituting Eq.~\eqref{eq:delta_n_result} into Eqs.~\eqref{Bcurrent} and \eqref{current2}
and expressing  the electric current and energy flux
densities in the form
\begin{equation}
\label{eq:response_matrix}
\left(
\begin{array}{c}
{\bf j}\\
{\bf j}_q
\end{array}
\right)=
\left(
\begin{array}{cc}
\hat{\sigma} & \hat{\beta} \\
\hat{\gamma} & \hat{\zeta}
\end{array}
\right)
\left(
\begin{array}{c}
{\bf E}\\
 \boldsymbol{\nabla} T
\end{array}
\right)
\end{equation}
we obtain the following expressions for the $zz$ components of the transport tensors
\begin{subequations}
\label{eq:results}
\begin{eqnarray}
\label{sigma}
\sigma_{zz} & = &
\left( \frac{e^{2} B}{h^{2}c} \right)^{2}
\ \int d \varepsilon \, \left( - \frac{d n_F(\varepsilon)}{d \varepsilon} \right) \frac{\tau_h(\varepsilon)}{\nu_{\alpha}(\varepsilon)}, \\
\label{kappa}
\beta_{zz}& = &  \left(\frac{e B}{h^2 c}\right)^{2}
\int d \varepsilon \, \frac{e (\varepsilon-\mu)}{T}\,   \frac{d n_F (\varepsilon)}{d \varepsilon} \frac{\tau_{eff}(\varepsilon)}{\nu_\alpha(\varepsilon)}, \\
\label{alpha}
\zeta_{zz} & = & \left(
\frac{e B}{h^{2}c}\right)^2
\int   d \varepsilon
 \frac{(\epsilon-\mu)^{2}}{T}
\frac{d n_F (\varepsilon)}{d \varepsilon}
 \frac{\tau_{eff}(\varepsilon)}{\nu_\alpha(\varepsilon)}.
\end{eqnarray}
\end{subequations}
By the Onsager symmetry principle $\gamma_{zz}= - \beta_{zz} T$. The electronic contribution to thermal conductivity $\kappa_{zz}$ may be expressed in terms of the electrical conductivity $\sigma_{zz}$ and other transport coefficients in Eq.~\eqref{eq:results} as~\cite{AbrikosovBook}
$\kappa_{zz}= -\zeta_{zz} - T\beta_{zz}^2/\sigma_{zz}$.
Since at high temperatures the considered  effects are small we concentrate on the  low temperature regime $T\ll \mu$. In this case  Eqs.~\eqref{eq:results} simplify to
\begin{subequations}
\label{eq:results_T0}
\begin{eqnarray}
\label{sigmaT0}
\sigma_{zz} (\mu) &= & \left(\frac{e^{2} B}{h^{2}c} \right)^{2}\frac{\tau_h (\mu)}{\nu(\mu)}, \\
\label{zetaEl}
\zeta_{zz} (\mu) &= &-  \frac{\pi^2 T}{3 e^2}\frac{\tau_{eff} (\mu)}{\tau_h(\mu)}  \, \sigma_{zz} (\mu), \\
\label{betaT0}
\beta_{zz} (\mu) & = &  e \, \frac{d \zeta_{zz} (\mu)}{ d\mu } .
\end{eqnarray}
\end{subequations}

Under the conditions specified above, the results in  Eq.~\eqref{eq:results_T0} are valid not only for Weyl and Dirac materials, but also for conventional conductors.  In the case of Weyl and Dirac semimetals  these equations  reproduce  results obtained in Refs.~\cite{SonSpivak2013,spivak2016magnetotransport}.
The difference between the conventional time- and centrosymmetric materials and Weyl semimetals is in value of the helicity relaxation time $\tau_h$.  In non-centrosymmetric Weyl semimetals with spin-nondegenerate electron spectrum the large value of $\tau_h/\tau_{tr}$ may be associated with the fact that for smooth disorder potential their inter-valley transitions associated with large momentum transfer are suppressed. In conventional conductors  the large value of  $\tau_h/\tau_{tr}$  arises from the large ratio of the Fermi energy to the band gap $E_{g}\ll \mu$, as described by Eq.~\eqref{timehierarchi}.
Taking $\nu (\mu)=\mu^{2}/\hbar^{3}u^{3}$   we get
\begin{equation}\label{sigmaDirac}
  \frac{\sigma_{zz}}{\sigma_D}\sim \left(\frac{ \hbar u eB}{ c \mu E_{g} }\right)^{2} \sim \left(\frac{ \hbar \omega_c}{  E_{g}} \right)^2.
\end{equation}
 Here $\sigma_{D}=  2  e^{2}\nu D$, with $D=u^{2}\tau_{tr}/3$ being the intra-valley diffusion coefficient, is the Drude conductivity,
 and $\omega_{c}\sim eB u/c \mu$ is the cyclotron frequency.
Equation~\eqref{sigmaDirac} may be considered as an upper  bound estimate for the magnitude of NLMR. The presence in the material of short range impurities, which can not be described by Eq.~\eqref{Dirac}, decreases the magnitude of the effect.

Of course there are other, ``conventional" contributions to the longitudinal magnetoresistance associated with the Fermi surface anisotropy (see for example Ref.~\onlinecite{Pal2010necessary} and references therein).
Typically at small magnetic field these contributions to magnetoconductivity scale as $(\sigma_{zz}(B)-\sigma(0))\sim \chi \sigma(0)(\omega_{c}\tau_{tr})^{2}$, and saturate  at $\omega_{c}\tau_{tr}\sim 1$.
Here $\chi<1$ is  a parameter characterizing the Fermi surface anisotropy.
Thus,  the condition for Eq.~\eqref{sigmaDirac} to dominate the longitudinal magneto-resistance is
\begin{equation}\label{LowHinequality}
 \chi (E_{g}\tau_{tr}/\hbar )^{2}<1.
\end{equation}
Even if this condition is not satisfied  the negative contribution to the magnetoresistance, Eq.~\eqref{sigmaDirac} can dominate at high magnetic fields where the conventional contribution saturates.  In this case the longitudinal magnetoresistance is a non-monotonic function of the magnetic field.  We note that a non-monotonic $B$-dependence of $\sigma_{zz}$ at low magnetic field  was observed in most of experiments on Dirac and Weyl metals.

In experiments on Dirac semimetals the observed magnetoconductance was a few times greater than the Drude value of the conductivity at ${\bf B}=0$. According to Eq.~\eqref{sigmaDirac} this may happen if $\hbar \omega_c /E_{g} >1$.
Note that in the quasi-classical limit $\hbar \omega_c \ll \mu$. Then  to have a big effect one should have $\mu \gg E_{g}$.

We  would like to point out  an important physical difference between expressions for the magnetoconductivity  $\sigma_{zz}$ in Eq.~\eqref{sigma} on the one hand, and $\kappa_{zz}$ and the thermoelectric $\alpha_{zz}$ in Eqs.~\eqref{kappa} and \eqref{alpha}   on the other hand.   The magnetoconductance in Eq.~\eqref{sigma} is controlled by the helicity relaxation time $\tau_h$, while  the magnetic field dependence of  the thermoelectric coefficient and thermal conductivity are controlled by $\tau_{eff}$, which is a combination of the helicity relaxation time $\tau_h$ and the  inelastic  relaxation time $\tau_{\epsilon}$. Thus,  according to Eq.~\eqref{zetaEl}  the Wiedemann-Franz law is violated at high temperatures where $\tau_{eff}\ll \tau$.   Furthermore, despite the conventional form of Eq.~\eqref{betaT0} the Mott relation also does not hold, $\beta_{zz} (\mu) \neq - \pi^2 T/(3 e) \partial_\mu \sigma_{zz} (\mu)$. The aforementioned difference and, consequently, the violation of the Wiedemann-Franz and Mott relations, can be traced to the difference in the  physical processes which determine magnetoconductance $\sigma_{zz}(B)$, and the magnetic field dependence of $\zeta_{zz}(B)$, and $\beta_{zz}(B)$. The magnetoconductance  is controlled by the long relaxation time $\tau_h$ of helicity imbalance at the Fermi level, which is created by acceleration of electrons in the lowest Landau level in the presence of the electric field.  This is similar to the chiral anomaly. As along as $T\ll \mu$ and $\tau_h$ slowly depends of the electron energy, the temperature dependence of the negative longitudinal magnetoresistance is weak. This explains why NLMR was observed up to relatively high temperatures.
In contrast, the temperature gradient does not create helicity imbalance, but only produces an energy imbalance between the electron populations with opposite helicity. The relaxation of the energy imbalance
is governed by the time $\tau_{eff}$, which at $\tau_h > \tau_{\epsilon}$ coincides by the inelastic relaxation rate, $\tau_{eff} \approx \tau_\epsilon$. As a result, the thermal conductivity and the thermoelectric coefficient exhibit a strong temperature dependence. In  the ``hydrodynamic" regime where $\tau_{tr} \gg \tau_{\epsilon}$ the described above contributions $\kappa_{zz}(B)$  and $\alpha_{zz}(B)$ become negligible compared  to the conventional contributions. Thus the dependence thermal conductivity and the thermoelectric coefficient on the magnetic field is unrelated to the chiral anomaly.

In conclusion, we have shown that positive contributions to the parallel magneto-conductance $\sigma_{zz}(B)$, the magnetic field dependent parallel thermal conductivity and
the thermoelectric coefficient $\beta_{zz}(B)$ can
exist not only in Weyl and Dirac semimetals, but also in
conventional cento-symmetric conductors as well.  We also would like to mention that the magnetic field dependence of the sound absorption coefficient exhibits similar properties \cite{spivak2016magnetotransport}.   We also expect that, similarly to the negative magnetoresistance of \emph{pn}-junctions in Weyl semimetals~\cite{li2016klein}, the magnetoresistance of \emph{pn}-junctions in Dirac semimetals with a sufficiently small gap $E_g$ will also be negative.

Our consideration focused on the quasiclassical regime $\hbar u/l_B  \ll \mu$.
In  the ultra-quantum limit, $\hbar u/l_B \gg\mu$, when only the zeroth Landau level is occupied
the situation is more complicated. In Weyl semimetals in the single particle approximation
an expression for conductivity in this regime was obtained in Ref.~\onlinecite{Nielsen1983},
$\sigma_{zz}\propto \frac{e^{2}u }{4\pi\hbar l_{B}^{2}}\, \tau_h$.
A similar result can be obtained for degenerate Dirac metals in the ultra-quantum regime.
The magnetic field dependence of the longitudinal magnetoresistance in this regime is controlled by the
corresponding magnetic field dependence of the helicity relaxation rate.
 The latter depends on the type of impurities. Its evaluation is not essentially different from the calculation of the backscattering rate in conventional semiconductors in the ultra-quantum limit.  In the context of conventional semiconductors in quantized magnetic field there is also an unrelated to the chiral anomaly mechanism of strongly anisotropic magnetoresistance, which may become negative in the longitudinal direction (see for example Refs.~\onlinecite{argyres1956longitudinal,Murzin2000electron,Goswami2015axial}).  It is related to the fact that in the presence of smooth potential in quantized magnetic field the small angle scattering is suppressed. As far as we know, this effect has never been observed in conventional semiconductors. An additional difficulty in interpreting magnetotransport measurements in the ultra-quantum regime is associated with the instability of the electron liquid with respect to charge density wave formation, which drives the system to the insulating state. In contrast, in the semiclassical limit, theoretical consideration of electron transport is free of aforementioned complications.

\acknowledgments{We thank E. Bettelheim for  useful discussions. The work of AVA was supported by the U.S. Department of Energy   Office of Science, Basic Energy Sciences under Award No. DE-FG02-07ER46452.}

\bibliography{biblioKERRupd,addrefs}
\bibliographystyle{apsrev}

\appendix

\section{\label{sec:helicity_relaxation} derivation of the helicity relaxation rate}

In this appendix we evaluate the helicity relaxation rate due to elastic scattering of electrons from impurities.
We are interested in the  regime where the energy of the electrons significantly exceeds the band gap, $\varepsilon \gg E_g$.
 In this case the helicity relaxation rate is parametrically smaller that the momentum relaxation rate and may be evaluated under the assumption of full momentum relaxation of the electron distribution with a given helicity.

The helicity relaxation rate due to elastic scattering of electrons from impurities is independent of the sign of the energy. Therefore, below we assume the energy to be positive. The electron states with momentum $\mathbf{p}$, positive energy $\epsilon_p= \sqrt{E_g^2 + u^2 p^2}$, and given helicity, $\alpha = \pm 1$ are described by the following wavefunctions
\begin{widetext}
\begin{equation}\label{eq:psi_+}
\Psi_+(\mathbf{p}) =
\left(
\begin{array}{c}
\frac{\epsilon_p + u p }{\sqrt{2 \epsilon_p (\epsilon_p + u p )}}\,  \cos (\theta/2)\\
\frac{\epsilon_p + u p }{\sqrt{2 \epsilon_p (\epsilon_p + u p )}}\, \sin (\theta/2) e^{i\phi}\\
\frac{E_g }{\sqrt{2 \epsilon_p (\epsilon_p + u p )}} \, \cos (\theta/2)\\
\frac{E_g }{\sqrt{2 \epsilon_p (\epsilon_p + u p )}} \,  \sin (\theta/2)  e^{i\phi}
\end{array}
\right) ,
\quad
\Psi_-(\mathbf{p}) =
\left(
\begin{array}{c}
- \frac{E_g }{\sqrt{2 \epsilon_p (\epsilon_p + u p )}} \, \sin (\theta/2)\\
\frac{E_g }{\sqrt{2 \epsilon_p (\epsilon_p + u p )}} \,  \cos (\theta/2)  e^{i\phi}\\
- \frac{\epsilon_p + u p }{\sqrt{2 \epsilon_p (\epsilon_p + u p )}} \,  \sin (\theta/2)\\
\frac{\epsilon_p + u p }{\sqrt{2 \epsilon_p (\epsilon_p + u p )}} \, \cos (\theta/2) e^{i\phi}
\end{array}
\right).
\end{equation}
\end{widetext}
Here we introduced the spherical angles $\theta$ and $\phi$ to define the direction of the electron momentum $\mathbf{p}$.

In the Born approximation the collision integral due to impurity scattering may be written in the form
\begin{eqnarray}
\label{eq:collision_integral}
I_\alpha\{n_\alpha (\mathbf{p}) \} & = &  \int d \mathbf{p}'  w (\mathbf{p}; \mathbf{p}') \left[  \frac{(\epsilon_p + u p)^2}{(2 \epsilon_p )^2}  \left( 1 + \frac{\mathbf{p} \cdot \mathbf{p}'}{p^2}\right) n_\alpha (\mathbf{p}') \right. \nonumber \\
&&\left. + \frac{E_g^2}{(2 \epsilon_p)^2}  \left( 1 - \frac{\mathbf{p} \cdot \mathbf{p}'}{p^2}\right) n_{- \alpha} (\mathbf{p}')  \right]  - \frac{n_\alpha (\mathbf{p})}{\tau}.
\end{eqnarray}
Here $n_{\pm \alpha} (\mathbf{p}') $ is the distribution function  of electrons with momentum $\mathbf{p}'$ and helicity $\pm \alpha $, $w (\mathbf{p}; \mathbf{p}') \propto \delta (p - p') \left|
\int d \mathbf{r} V(\mathbf{r}) e^{i (\mathbf{p} - \mathbf{p}')\cdot \mathbf{r}}
\right|^2 $,  and the expression in the square brackets describes the square of the overlap of spinor amplitudes \eqref{eq:psi_+} in states $\mathbf{p}$ and $\mathbf{p}'$  with the same (first term) and opposite (second term) helicities. Finally,
 the ``out" relaxation rate is given by
\begin{eqnarray}
\label{eq:tau}
\frac{1}{\tau} & = &  \int d \mathbf{p}'  w (\mathbf{p}; \mathbf{p}') \left[ \frac{(\epsilon_p + u p)^2}{(2 \epsilon_p )^2} \left( 1 + \frac{\mathbf{p} \cdot \mathbf{p}'}{p^2}\right) \right. \nonumber \\
&&\left.    +\frac{E_g^2}{(2 \epsilon_p)^2}  \left( 1 - \frac{\mathbf{p} \cdot \mathbf{p}'}{p^2}\right)   \right].
\end{eqnarray}

In the approximation of full momentum relaxation the distribution function depends only the the energy $\varepsilon$ and helicity $\alpha$ of the electrons, and the collision integral simplifies to
\begin{equation}
\label{eq:collision_integral_helicity}
I_\alpha \{ n(\varepsilon)\} = - \frac{n_\alpha (\varepsilon)  - n_{-\alpha} (\varepsilon)}{\tau_h (\varepsilon)},
\end{equation}
where $\tau_h$ is the helicity relaxation time given by
\begin{equation}
\label{eq:tau_h}
\frac{1}{\tau_h (\varepsilon)}=   \frac{E_g^2}{4 \varepsilon^2} \,  \int d \mathbf{p}'  w (\mathbf{p}; \mathbf{p}')   \left( 1 - \frac{\mathbf{p} \cdot \mathbf{p}'}{p^2}\right)  .
\end{equation}

Note that in the limit $E_g/\epsilon_p \to 0$ the transport scattering rate is given by
\begin{equation}
\label{eq:tau_tr}
\frac{1}{\tau_{tr} (\varepsilon)} =   \int d \mathbf{p}'  w (\mathbf{p}; \mathbf{p}') \left( 1 + \frac{\mathbf{p} \cdot \mathbf{p}'}{p^2}\right)  \left( 1 - \frac{\mathbf{p} \cdot \mathbf{p}'}{p^2}\right).
\end{equation}

Therefore the helicity relaxation rate may be expressed in terms of the transport relaxation rate in the form of Eq.~\eqref{timehierarchi}, where $\xi$ is a factor of order unity given by
\begin{equation}
\label{eq:Upsilon}
\xi = \frac{   \int d \mathbf{p}'  w (\mathbf{p}; \mathbf{p}') \left( 1 + \frac{\mathbf{p} \cdot \mathbf{p}'}{p^2}\right)  \left( 1 - \frac{\mathbf{p} \cdot \mathbf{p}'}{p^2}\right)}{\int d \mathbf{p}'  w (\mathbf{p}; \mathbf{p}')   \left( 1 - \frac{\mathbf{p} \cdot \mathbf{p}'}{p^2}\right) }.
\end{equation}
In the limiting case of small angle scattering $\xi =2$, while in for  point-like impurities $\xi=2/3$.

\end{document}